\documentclass[prl, twocolumn,preprintnumbers,aps]{revtex4}
\usepackage{graphicx}
\usepackage{makeidx}

\usepackage{color}

\newcommand{\be}{\begin{eqnarray}}
\newcommand{\ee}{\end{eqnarray}}

\begin{document}
\title{Modified Entropic Gravitation in Superconductors}

\author{Clovis Jacinto de Matos$^{\rm 1}$}

\affiliation{$^{\rm 1}$European Space Agency, 8-10 rue Mario
Nikis, 75015 Paris, France}

\date{12 August 2011}

\preprint{}

\begin{abstract}
Verlinde recently developed a theoretical account of gravitation
in terms of an entropic force. The central element in Verlinde's
derivation is information and its relation with entropy through
the holographic principle. The application of this approach to the
case of superconductors requires to take into account that
information associated with superconductor's quantum vacuum energy
is not stored on Planck size surface elements, but in four volume
cells with Planck-Einstein size. This has profound consequences on
the type of gravitational force generated by the quantum vacuum
condensate in superconductors, which is closely related with the
cosmological repulsive acceleration responsible for the
accelerated expansion of the Universe. Remarkably this new
gravitational type force depends on the level of breaking of the
weak equivalence principle for cooper pairs in a given
superconducting material, which was previously derived by the
author starting from similar principles. It is also shown that
this new gravitational force can be interpreted as a surface
force. The experimental detection of this new repulsive
gravitational-type force appears to be challenging.
\end{abstract}

\maketitle

{\sc Introduction---}Recently Verlinde introduced the interesting
possibility of gravitation being an entropic force
\cite{Verlinde}. This approach to the physical nature of
gravitation depends  strongly on the way in which information is
stored in physical systems. The scale at which the information is
stored and its location over a surface or a volume are fundamental
to determine the type of gravitational force generated by physical
systems. Thus the application of Verlinde's procedure to derive
the gravitational force produced by a physical system needs to
take into account the physical nature of the vacuum associated to
its lowest energy level. In the following the electromagnetic
zero-point model for the vacuum in superconductors, developed by
Beck, Mackey and the author, together with the quantization of
information over four-dimensional Planck-Einstein size cells,
required by this model, are taken into consideration when applying
Verlinde's procedure to deduce the entropic gravitational force
generated by the quantum vacuum present in a superconductor. It
appears that the quantum vacuum in superconductors generates a
gravitational type repulsive force, which is proportional to the
superconductor's volume and depends on the level of breaking of
the weak equivalence principle for Cooper pairs (for a given
superconductor). Following its derivation, The discussion
indicates that one can also deduce this force from Sivaram's
hydrodynamic procedure to derive Newton gravitational law assuming
masses immersed in a cosmic sea of vacuum energy. This exercise
shows that one can interpret our basic result, Equ.(\ref{E3}), not
only as an entropic force but also in terms of a surface force.
The comparison is also made with the cosmological repulsive force,
and with a similar force which could arise if we consider that the
information of the universe is stored in four dimensional Planck
size cells, instead of being stored on Planck size area elements
as stipulated by Susskind holographic principle. Although
challenging to detect experimentally, in the conclusion one
discusses some experimental possibilities to detect the new type
of gravitational repulsive force proposed in the present paper,
and we close by introducing some theoretical avenues for future
work.

{\sc Information Theory and Universal Scaling Relations for the
Mass and Size of Physical Systems---} Bekenstein \cite{Beken}
demonstrated from quantum statistical physics and thermodynamics
that the ratio of entropy $S$ to mean energy $E_0$ of a spherical
system in its rest frame, is directly proportional to its
effective radius $R$.
\begin{equation}
\frac{S}{E_0}=2\pi \frac{k}{\hbar c} R\label{I0}
\end{equation}
Where $k$ is Boltzmann constant, $c$ is the speed of light in
vacuum, and $\hbar$ is Planck's constant divided by $2\pi$. The
theory of information establishes the general result that the
maximum entropy of a given system is directly proportional to the
amount of information $I$, in bits, that the system can store
\cite{Brillouin}.
\begin{equation}
S=\ln(2) k I \label{I1}
\end{equation}
Substituting Equ.(\ref{I1}) in Equ.(\ref{I0}) one obtains the
Bekenstein bound \cite{Bekenstein}, which defines the maximum
storage capacity of a system in function of its mean energy $E_0$
and radius $R$.
\begin{equation}
I\leq\frac{2\pi}{\ln2}\frac{1}{\hbar c} E_0 R\label{I2}
\end{equation}
One can also express this result in function of the system's
proper mass $M_0=E_0/c^2$.
\begin{equation}
I\leq\frac{2\pi}{\ln2} \frac{c}{\hbar}M_0 R\label{I2}
\end{equation}

Susskind further proposed the holographic principle, which states
that the information contained in a physical system is stored on
its boundary, the total number of bits being proportional to the
area $A$, which circumscribes the system \cite{Susskind}.
\begin{equation}
I=\frac{A}{l_p^2}\label{I3}
\end{equation}
where $l_p=\sqrt{G\hbar/c^3}$ is Planck's length. From the
equality between Equ.(\ref{I2}) and Equ.(\ref{I3}) one deduces
Sivaram's scaling relation \cite{Sivaram} for the mass and size of
the Universe, black holes and more generally of objects trapped in
their own gravitational field.
\begin{equation}
\frac{M}{R}\sim\frac{c^2}{G}\label{I4}
\end{equation}

In previous work the author argued that Cooper pairs in
superconductors could break the weak equivalence principle on the
basis that the information $I'$ associated with the Cooper pair
condensate is stored in four dimensional Planck-Einstein size
cells \cite{cloclo}.
\begin{equation}
I'= \frac {V}{l_{pe}^4}\label{I5}
\end{equation}
Where $V$ is the spherical four-volume circumscribing the
superconductor, and $l_{pe}=(\hbar G/c^3\Lambda)^{1/4}$ is the
Planck-Einstein length, which is a typical scale for dark energy
when explained through the cosmological constant $\Lambda$
\cite{Beck1}. Since all forms of matter and energy in the universe
have approximately the same age as the Universe:
$T_U=1/c\Lambda^{1/2}$, the four volume of any physical system is
directly proportional to its spatial volume. If we consider a
spherical system with radius $R$ we have:
\begin{equation}
V=\frac{4}{3} \pi R^3 c T_U =\frac{4}{3} \pi R^3 \Lambda^{- 1/2}
\label{I6}
\end{equation}
Substituting Equ.(\ref{I6}) in Equ.(\ref{I5}) one obtains:
\begin{equation}
I'=\frac{4\pi}{3}\frac{c^3\Lambda^{1/2}}{\hbar G} R^3\label{I7}
\end{equation}
From the equality between Bekenstein's bound, Equ.(\ref{I2}), and
Equ.(\ref{I7}) one deduces Sivaram's scaling relation
\cite{Sivaram} for the mass and size of objects with low surface
gravity.
\begin{equation}
\frac{M}{R^2}\sim\frac{c^2}{G}\sqrt{\Lambda}\sim 1\label{I8}
\end{equation}

The fact that different versions of the holographic principle,
Equ.(\ref{I3}) and Equ.(\ref{I5}), lead to meaningful scaling
relations for mass and size of physical objects, when compared to
the Bekenstein bound, is an interesting result on its own right.
It also demonstrates the great generality of Bekenstein's result,
Equ.(\ref{I2}), and is encouraging the further use of the
four-dimensional "holographic-type" relation, Equ.(\ref{I7}) in
the following part of the present work.

{\sc Classical Entropic Gravitation---} Let us consider a body
with mass $M_0$ circumscribed by a spherical boundary with radius
$R$, which can be larger than the physical size of the body which
it contains, and a probing mass $m_0<<M_0$ which is approaching
this spherical boundary.

Verlinde derives Newton's gravitational force as an entropic force
from the second law of thermodynamics.
\begin{equation}
F\Delta R= T \Delta S\label{C0}
\end{equation}
Where $F$ is the gravitational force, $\Delta R$ is an
infinitesimal variation of the radius of the spherical boundary
circumscribing $M_0$, $T$ is the equilibrium temperature of the
spherical boundary, and $\Delta S$ is the variation of entropy
associated with the variation of the information stored on the
spherical boundary around $M_0$ resulting from $m_0$ approaching
this boundary.

From the Bekenstein relation between the entropy and the energy of
a system, Equ.(\ref{I0}), one deduces that the variation of
entropy related to the information on the boundary of the system
is:
\begin{equation}
\Delta S=2\pi \frac{k}{\hbar c} \Delta R \ E_0\label{C1}
\end{equation}
Verlinde's considers that this entropy change is caused by the
particle of mass $m_0=E_0/c^2$ approaching a part of the sphere
circumscribing the mass $M_0$. Thus one obtains:
\begin{equation}
\Delta S=2\pi k \frac{m_0 c}{\hbar} \Delta R \label{C2}
\end{equation}
One can understand this entropy variation as the addition of the
entropy carried out by the spherical boundary with radius $\Delta
R$ circumscribing the mass $m_0$, to the entropy stored on the
spherical boundary circumscribing the mass $M_0$.

The entropic approach to gravitation, Equ.(\ref{C0}),  requires to
have a temperature $T$ in order to have a force. In Verlinde's
model this temperature corresponds to the equilibrium temperature
of the boundary circumscribing the mass $M_0$. This temperature is
calculated assuming that the the energy $M_0 c^2$ is equally
distributed among the $I$ bits available over the spherical
surface $A=4\pi R^2$.
\begin{equation}
M_0c^2=\frac{1}{2}I k T\label{C3}
\end{equation}
Substituting Equ.({\ref{I3}) in Equ.(\ref{C3}), one obtains:
\begin{equation}
T=\frac{G\hbar}{2\pi k c} \frac{M}{R^2}\label{C4}
\end{equation}
Substituting Equ.(\ref{C4}), and Equ.(\ref{C2}) in Equ.(\ref{C0})
one obtains Newton's law of gravitation.
\begin{equation}
F=G\frac{Mm}{R^2}=6.67\times 10^{-11} \frac{Mm}{R^2}\label{C5}
\end{equation}

{\sc Entropic Gravitation in Superconductors---} Beck, Mackey, and
the author \cite{Beck} \cite{dematos} have developed an
electromagnetic model of vacuum energy in superconductors,
starting from the assumption that the virtual photons with
zero-point energy
\begin{equation}
\epsilon=h\nu/2\label{eq1}
\end{equation}
(where $h$ is Planck's constant) form a condensate below the
superconductor's critical temperature $T_c$ with energy density:
\begin{equation}
\rho^*=\int_0^{\nu_c}\frac{1}{2}h\nu\frac{4\pi}{c^3}\nu^2d\nu.\label{eq2}
\end{equation}
In Eq(\ref{eq2}) the two possible polarization states of the
photon are considered, and $\nu_c$ is a certain maximum cutoff
frequency. This frequency is calculated by assimilating the
condensate of virtual photons with energy $\epsilon_c= h\nu_c/2$
with a black body thermal bath of ordinary photons at the
superconductor's critical temperature $T_c$ with mean energy
$\bar{\epsilon_c}=h\nu_c/e^{h\nu/kT_c}-1\label{eq3}$,
\begin{equation}
\frac{1}{2} h\nu_c=\frac{h\nu_c}{e^{\frac{h\nu}{kT_c}}-1}.
\label{eq4}
\end{equation}
This condition is equivalent to:
\begin{equation}
h\nu_c=\ln3kT_c\label{eq5}
\end{equation}
Substitution of eq.(\ref{eq5}) in eq.(\ref{eq2}) leads to the law
defining the density of the electromagnetic zero-point energy
condensate in function of the superconductor's critical
temperature, $T_c$.
\begin{equation}
\rho^*=\frac{\pi \ln^4 3}{2}\frac{k^4}{(ch)^3}T_c^4\label{eq6}
\end{equation}

A non-vanishing cosmological constant can be interpreted in terms
of a non-vanishing vacuum energy density, $\rho_0$, associated
with a cosmological quantum vacuum field.
\begin{equation}
\rho_0=\frac{c^4}{8 \pi G} \Lambda\sim 10^{-29} g~cm^{-3} \simeq
3.88 e V/mm^3  \label{eq23}
\end{equation}
where $\Lambda=1.29\times 10^{-52} [m^{-2}]$ is the cosmological
constant \cite{Spergel}.

In \cite{cloclo}, the author argued that the E\"{o}tv\"{o}s factor
$\chi$ quantifying the level of breaking of the weak equivalence
principle by Cooper pairs in superconductors, is proportional to
the ratio of the density of electromagnetic zero-point energy
(\ref{eq6}) to the density of cosmological vacuum energy
(\ref{eq23}).
\begin{equation}
\chi=\frac{\Delta
m_i}{m_g}=\frac{3}{2}\frac{\rho^*}{\rho_0}=\frac{3\ln^4 3}{8
\pi}\frac{k^4G}{c^7\hbar^3 \Lambda} T_c^4\label{12}.
\end{equation}
Where $\Delta m_i$ is the anomalous Cooper pair inertial mass
excess, which has been measured by Tate and Cabrera for the case
of superconducting Niobium \cite{Tate01}, and $m_g$ is the Cooper
pair's gravitational rest mass, which is assumed to be equal to
the Cooper pairs classical bare mass. Remarkably, Equ.(\ref{12})
connects the five fundamental constants of nature $k,G,c,\hbar,
\Lambda$ with measurable quantities in a superconductor, $\chi$
and $T_c$.We may define a Planck-Einstein temperature scale
$T_{PE}$ streamlined with the Planck-Einstein length introduced
above through Equ.(\ref{I5}).
\begin{equation}
T_{PE}=\frac{1}{k}\Bigg(\frac{c^7\hbar^3
\Lambda}{G}\Bigg)^{1/4}=60.71 K. \label{13}
\end{equation}
Equ.(\ref{12}) can then be written in a more elegant form
\cite{dematos}.
\begin{equation}
\chi=\frac{3\ln^4 3}{8
\pi}\Bigg(\frac{T_c}{T_{PE}}\Bigg)^4.\label{14}
\end{equation}

We have now all the elements to apply Verlinde's entropic model of
gravity to the case of superconductor's electromagnetic zero-point
energy condensate: Like in the previous section, one starts
considering that a spherical superconductor with mass $M_0$ and
volume $V_0$ is circumscribed by a spherical boundary of radius
$R$, in the neighborhood of which a probing mass $m_0$ is moving.
The variation of entropy caused by the mass $m_0$ is the same as
previously calculated for classical materials, i.e.
Equ.(\ref{C2}). However the information related with the
electromagnetic zero-point energy present in the superconductor is
equally stored between the $I'$ Planck-Einstein cells,
Equ.(\ref{I5}), available over the four-volume $V=\frac{4}{3} \pi
R^3\Lambda^{-1/2}$ surrounding the superconductor.
\begin{equation}
\rho^* V_{sc}=\frac{1}{2} k T \frac{V}{l_{pe}^4}\label{E0}
\end{equation}
Where $\rho^*$ is the density of electromagnetic zero-point energy
in the superconductor, Equ.(\ref{eq6}), $V_{sc}=\frac{4}{3} \pi
R_{sc}^3$ is the superconductor volume, and $T$ is the equilibrium
temperature of the four-volume $V$ (including, of course, its
spherical spatial boundary). Substituting Equ.(\ref{eq6}), and the
Planck-Einstein length $l_{pe}=(\hbar G/c^3\Lambda)^{1/4}$ in
Equ.(\ref{E0}) one obtains for the temperature:
\begin{equation}
T=\frac{(\ln3)^4}{2}\frac{k^3 G}{h^2
c^6\Lambda^{1/2}}T_c^4\Big(\frac { R_{sc}}{R}\Big)^3\label{E1}
\end{equation}
Substituting Equ.(\ref{C2}) and Equ.(\ref{E1}) in Equ.(\ref{C0})
one obtains the gravitational type force $F_{zp}$ produced by the
superconductor's electromagnetic zero-point energy condensate.
\begin{equation}
F_{zp}=\frac{(\ln3)^4}{4\pi}\frac{k^4G}{\hbar^3 c^5 \Lambda^{1/2}}
T_c^4 \Big(\frac{R_{sc}}{R}\Big)^3 m\label{E2}
\end{equation}
Substituting the Cooper pair's E\"{o}tv\"{o}s factor $\chi$
Equ.(\ref{12}) in Equ.(\ref{E2}), I get my basic result in a
synthetic form.
\begin{equation}
F_{zp}=\frac{1}{3} c^2 \Lambda^{1/2} \ \frac {m_0 \ \chi R_{sc}^3
}{R^3}\sim 3.4\times10^{-10}  \ \frac {m \ \chi R_{sc}^3
}{R^3}\label{E3}
\end{equation}
For the case of Niobium $\chi\sim9.35\times10^{-5}$, substituting
this value in Equ.(\ref{E3}) and dividing by $m_0$ one gets the
gravitational acceleration generated by the quantum vacuum in
superconducting Niobium:
\begin{equation}
a_{zp}\sim 3.18\times10^{-14}  \ \frac {\ R_{sc}^3
}{R^3}\label{E4}
\end{equation}

The dependence on the inverse of the cube of the distance with
respect to the superconductor, indicates that the zero-point
gravitational type force, Equ.(\ref{E3}) decreases faster than the
Newtonian gravitational force which is proportional to the inverse
of the square of the distance from the source. Thus at large
distances from the superconductor's center of mass, Newtonian
gravity will largely dominate over the gravitational zero point
force Equ.(\ref{E3}).

{\sc Discussion---} Comparing the gravitational zero-point force
of a superconductor, Equ.(\ref{E3}), with the cosmological
repulsive force\cite{Ohanian}, $F_c$, responsible for the observed
accelerated expansion of the Universe.
\begin{equation}
F_c=\frac{1}{3} c^2 \Lambda m r\label{D0}
\end{equation}
one sees that both forces have similar expressions, and that both
should be repulsive forces since latest cosmological observations
reveal that the cosmological constant is positive \cite{Spergel},
$\Lambda > 0$.

The total gravitational force produced by a superconductor should
be understood has being the sum of the classical Newtonian
attractive gravitational force originating form the total
gravitational mass of the superconductor $M_{sc}$, and the
repulsive zero-point gravitational force originating from the
Electromagnetic zero-point condensate in the superconductor.
\begin{equation}
\vec{F}_{grav}=\frac{m}{R^2}\Big(\frac{1}{3} c^2 \Lambda ^{1/2}
\frac{\chi R_{sc}^3}{R}\ -\ G M_{sc}\Big) \hat{r}\label{D1}
\end{equation}
where $\hat{r}$ is the radial unit vector pointing outwards with
respect to the superconductor center of mass.

In the context of the results just derived, it is instructive to
compare Verlinde's procedure to derive the Newtonian gravitational
force with Sivaram's derivation, which achieves the same result on
the basis of the laws of hydrodynamics \cite{Sivaram2}. It is
impressive to see that by applying Sivaram's procedure to the case
of the electromagnetic zero-point model of vacuum in
superconductors, introduced above, one succeeds to find back the
same repulsive zero-point gravitational force obtained in
Equ.(\ref{E3}).

Sivaram starts by considering that a mass $M_0$, circumscribed by
a spherical shell with radius $R$, will affect the density of the
vacuum energy due the the spacetime curvature it generates.
\begin{equation}
P'_{vac}=P_{vac}\Big(1-\frac{2GM_0}{Rc^2}\Big)\label{D2}
\end{equation}
where $P'_{vac}$ vacuum pressure (or identically, the vacuum
energy density) inside the shell, $P_{vac}=\rho_0=c^4\Lambda/8 \pi
G$ is the ambient vacuum pressure outside the shell. Thus the
pressure difference across the shell will be.
\begin{equation}
\Delta P_{vac}=P_{vac} - P'_{vac}=P_{vac}\frac{GM}{c^2R}
\label{D3}
\end{equation}
Inspired by the Archimedes principle, Sivaram proposed that any
object of mass $m_0$ in the vacuum fluid displaces a volume $V$,
such that:
\begin{equation}
m_0 c^2= P_{vac} V \label{D4}
\end{equation}
Multiplying both sides of Equ.(\ref{D3}) by the displaced volume
of vacuum fluid $V$, one obtains an expression analogous to the
one giving the buoyant Archimedes force.
\begin{equation}
\Delta P_{vac}V=P_{vac}V\frac{GM}{c^2R}\label{D5}
\end{equation}
Substituting Equ.(\ref{D4}) in Equ.(\ref{D5}) one obtains the
Newtonian gravitational potential energy, which is binding
together the masses $M_0$ and $m_0$.
\begin{equation}
U_g=\frac{GM_0 m_0}{R}\label{D6}
\end{equation}
Taking the gradient of Equ.(\ref{D6}), i.e. the pressure gradient
of the vacuum energy, one obtains Newton's law.
\begin{equation}
\vec F = - G \frac{M_0 m_0}{R^2} \hat{R}\label{D7}
\end{equation}
Thus we have a hydrodynamics derivation of Newton's law indicating
that one can interpret Newton's gravitational force as a Buoyant
type force between masses immersed in a sea of vacuum energy.

To apply this procedure to the case of a superconductor one must
take into account that the vacuum energy density inside a
superconductor $\rho^*$ is given by Equ.(\ref{eq6}).
\begin{equation}
\rho^*=\frac{2}{3} \chi \rho_0=\frac{2}{3} \chi P_{vac} \label{D8}
\end{equation}
If one imagines a spherical membrane with radius $R$
circumscribing the spherical superconductor of radius $R_{sc}$,
the vacuum pressure inside the membrane $P'_{vac}$ is:
\begin{equation}
P'_{vac}=P_{vac} - \rho^*\label{D9}
\end{equation}
Substituting Equ.(\ref{D8}) into Equ.(\ref{D9}) one obtains the
pressure difference across the membrane.
\begin{equation}
\Delta P_{vac}= \frac{2}{3} \chi P_{vac}\label{D10}
\end{equation}
Setting the total vacuum energy in the superconductor,
$\frac{4}{3}\pi R^3_{sc} \rho^*$, equal to the total difference of
vacuum energy across the membrane of radius $R$.
\begin{equation}
\Delta P_{vac}\frac{4}{3} \pi R^3= \frac{2}{3} \chi P_{vac}
\frac{4}{3}\pi R^3_{sc} \label{D11}
\end{equation}
One obtains the pressure difference across the membrane in
function of the inverse of the cube of the distance from the
superconductor.
\begin{equation}
\Delta P_{vac}= \frac{2}{3} \chi P_{vac}\Big(\frac
{R_{sc}}{R}\Big)^3\label{D12}
\end{equation}
Sivaram \cite{Sivaram} scaling relation Equ.(\ref{I8}) also
implies a universal relation for the surface tension $T$
(expressed in $Joules/Area$) of physical systems, which states
that the surface tension of a physical system is always
approximately equal to the surface tension of the Universe.
\begin{equation}
T=\rho_0
R_{Universe}=\frac{\rho_0}{\Lambda^{1/2}}=\frac{P_{vac}}{\Lambda
^{1/2}}\label{D13}
\end{equation}
Substituting Equ.(\ref{eq23}) into Equ.(\ref{D13}) one gets:
\begin{equation}
T\sim \frac{\Lambda ^{1/2} c^4}{G}\label{D14}
\end{equation}
Applying Sivaram's universal scaling relation for tension to the
case of a probing mass $m_0$ moving in the neighborhood of the
mass $M_0$, one can calculate the cross sectional area $\sigma$ of
the moving particle $m_0$, which defines the gravitational surface
interaction between the two masses. This is achieved by
substituting the tension $T=m_0c^2/\sigma$ in Equ.(\ref{D13}).
\begin{equation}
\frac{mc^2}{\sigma}=\frac{P_{vac}}{\Lambda ^{1/2}}\label{D15}
\end{equation}
Multiplying both sides of Equ.(\ref{D12}) by the area $\sigma$
calculated from Equ.(\ref{D15}), one obtains a surface
gravitational type force very close to the zero point
gravitational type force, Equ.(\ref{E3}), derived from Verlinde's
procedure applied to superconductors.
\begin{equation}
F_s=\frac{2}{3} c^2 \Lambda ^{1/2} \frac{m_0 \chi
R_{sc}}{R^3}\label{D16}
\end{equation}
Note that since $P_{vac}<P'_{vac}$ this force is repulsive with
respect to the superconductor since it will always point from high
to low vacuum pressure. Thus one should also observe a tiny
negative pressure on the superconductor \cite{cloclo}. The
important lesson one learns from this derivation is that the
repulsive zero-point gravitational type force produced by
superconductors can be interpreted not only in terms of an
entropic force but also as a surface-type force.

Before one closes the discussion, let us deduce the consequences
of Sorkin's alternative to the holographic principle to account
for the small value of the cosmological constant \cite{Beck1}.
Sorkin's suggested that the total amount of information stored in
the Universe is directly proportional to the universe four-volume
\cite{Sorkin}\cite{Maqbool}:
\begin{equation}
I''=\frac{V}{l_p^4}\label{DD2}
\end{equation}
Where $V=\frac{4}{3} \pi R^3 c^{-1}\Lambda^{-1/2}$ is the four
volume of the universe and $l_p=\sqrt{G\hbar/c^3}$ is Planck's
length. From the equality between Equ.(\ref{DD2}) and the
Bekenstein bound Equ.(\ref{I2}) one obtains a new scaling relation
for mass and length of physical bodies:
\begin{equation}
\frac{M}{R^2}= \frac{2}{3} \ln 2 \frac {c^5}{c^2 \hbar
\Lambda^{1/2}}\label{DD3}
\end{equation}
Calculating the equilibrium temperature of the universe spherical
boundary following Verlinde's procedure, already used several
times above,
\begin{equation}
M_0 c^2= \frac{1}{2}kT I'' \label{DD4}
\end{equation}
and substituting Equ.(\ref{DD2}) in Equ.(\ref{DD4}) one obtains
for the equilibrium temperature:
\begin{equation}
T=\frac{3}{2\pi} \frac{\Lambda^{1/2} G^2 \hbar^2}{kc^4} \frac
{M}{R^3}\label{DD5}
\end{equation}
Substituting Equ.(\ref{DD5}) and Equ.(\ref{C2}) in Equ.(\ref{C0})
one obtains the gravitational force law:
\begin{equation}
F=3 \frac{\Lambda ^{1/2} G^2 \hbar}{c^3} \frac {M_0 m_0}{R^3}\sim
5.93\times 10^{-106} \frac {M_0 m_0}{R^3}\label{DD6}
\end{equation}
Comparing relations Equ.(\ref{DD3}) and Equ.(\ref{DD6}) with the
equivalent relations Equ.(\ref{I4}), Equ.(\ref{C5}) and
Equ.(\ref{I8}), Equ.(\ref{E3}) respectively, one sees that
relations Equ.(\ref{DD3}) and Equ.(\ref{DD6}) cannot be valid for
physical systems inside our own Universe. One could speculatively
propose that the scaling relation Equ.(\ref{DD3}) could be
understood as holding for possible different universes, and that
the force law Equ.(\ref{DD6}) could be related with the
gravitational force between different universes of total mass
$m_0$ and $M_0$ respectively. This also contributes to indicate
that our basic result Equ.(\ref{E3}) can only be obtained in the
context of the Planck-Einstein scale.

{\sc Conclusions---} Combining Verlinde's entropic account of
gravitation, Equ.(\ref{C0}), with the electromagnetic zero-point
model of quantum vacuum in superconductors, Equ.(\ref{eq6}), one
deduces a new type of repulsive gravitational force generated by
the electromagnetic zero-point vacuum energy condensate contained
in the crystal of a superconductor Equ.(\ref{E3}). Comparing this
derivation with Sivaram's procedure to obtain Newton's
gravitational law from the hydrodynamics of vacuum, applied to
superconductors, one deduces that this repulsive zero-point
gravitational force is also a surface force. The results and
discussion presented in this paper, suggests carrying out
E\"{o}tv\"{o}s type experiments with superconducting masses aiming
at measuring any anomalous value of the universal gravitational
constant, which could be accounted for by the repulsive
gravitational force Equ.(\ref{E3}). In addition, one could also
suggest repeating the small scale tests of the gravitational
inverse square law as performed by Adelberger et al but with
superconducting probing masses, instead of classical masses
\cite{Beck1}, in order to confirm the dependence on the inverse of
the cube of the distance of the new gravitational repulsive law
Equ.(\ref{E3}). On the theoretical side it would be interesting to
explore any possible relationship between the surface force
derived in the present work, Equ.(\ref{E3}), with the Tao surface
force in superconducting millimetric balls \cite{Tao}, and with
the surface force predicted by Ulf Leonhardt for the optical
analogue of the Iordanskii force in rotating Bose-Einstein
condensates \cite{Ulf}.

{\sc Acknowledgments---} The author would like to acknowledge Mr.
Luca del Monte for calling his attention to Verlinde's work on
entropic gravitation, and for stressing the relevance of this
thermodynamic approach to the nature of gravitation for the
author's own research.

\end{document}